\documentclass[10pt,conference]{IEEEtran}
\usepackage{epsfig,setspace,amsmath,epsf,amssymb,bm,theorem,cite,graphicx, epstopdf,algorithm,algpseudocode,float,color,mathtools,authblk}
\usepackage[table,xcdraw]{xcolor}
\usepackage{booktabs}
\usepackage{subfigure}

\IEEEoverridecommandlockouts
\allowdisplaybreaks

\begin{document}

\title{Modeling AoII in Push- and Pull-Based Sampling of Continuous Time Markov Chains}

\author[1]{Ismail Cosandal}
\author[2]{Nail Akar}
\author[1]{Sennur Ulukus}

\affil[1]{\normalsize University of Maryland, College Park, MD, USA}
\affil[2]{\normalsize Bilkent University, Ankara, T\"{u}rkiye}

\maketitle

\let\thefootnote\relax\footnotetext{This work is done when N.~Akar is on sabbatical leave as a visiting professor at University of Maryland, MD, USA, which is supported in part by the Scientific and Technological Research Council of T\"{u}rkiye  (T\"{u}bitak) 2219-International Postdoctoral Research Fellowship Program.}

\begin{abstract}
Age of incorrect information (AoII) has recently been proposed as an alternative to existing information freshness metrics for real-time sampling and estimation problems involving information sources that are tracked by remote monitors. Different from existing metrics, AoII penalizes the incorrect information by increasing linearly with time as long as the source and the monitor are de-synchronized, and is reset when they are synchronized back. While AoII has generally been investigated for discrete time information sources, we develop a novel analytical model in this paper for push- and pull-based sampling and transmission of a continuous time Markov chain (CTMC) process. In the pull-based model, the sensor starts transmitting information on the observed CTMC only when a pull request from the monitor is received. On the other hand, in the push-based scenario, the sensor, being aware of the AoII process, samples and transmits when the AoII process exceeds a random threshold. The proposed analytical model for both scenarios is based on the construction of a discrete time MC (DTMC) making state transitions at the embedded epochs of synchronization points, using the theory of absorbing CTMCs, and in particular phase-type distributions. For a given sampling policy, analytical models to obtain the mean AoII and the average sampling rate are developed. Numerical results are presented to validate the analytical models as well as to provide insights on optimal sampling policies under sampling rate constraints. 
\end{abstract}

\section{Introduction}

In real-time sampling and estimation problems involving information sources (or sensors) that are tracked by remote monitors, the performance of the estimator is indirectly quantified by a number of information freshness metrics, such as age of information (AoI) \cite{yates2019,yates-survey} and metrics derived from it, e.g., age of incorrect information (AoII) \cite{maatouk2020}, query-age of information \cite{chiariotti2022query}, age penalty \cite{champati2022detecting}, etc. The most common feature of these metrics is that the penalty linearly increases with time while the information is stale at the monitor. In particular, AoII has been proposed in \cite{maatouk2020} to quantify information freshness by allowing AoII to increase only when there is a mismatch between the observed source and its estimator, and it can be reset back to zero even when this agreement is satisfied without having to take a new sample. It also differs from other metrics such as binary freshness \cite{akar2023optimum} and MMSE (minimum mean square error) by further increasing the penalty when the estimation is kept incorrect for a long time.

AoII metric has generally been used in discrete time settings \cite{maatouk2020,kam2020age,kriouile2021minimizing,kriouile2022pull,chen2021minimizing}, and a system model in continuous time is studied for the first time in this paper, to the best of our knowledge. Most practical systems are of continuous time nature which is the main reason for our choice of a continuous time setting. In \cite{maatouk2020,kam2020age,chen2021minimizing}, push-based transmission models are investigated, and an optimum threshold policy for a discrete time Markov chain (DTMC) source with homogeneous transition probabilities is proposed by using the Markov decision processes (MDP).
In \cite{kriouile2022pull}, a pull-based transmission system for AoII is studied in which a monitor observes multiple homogeneous Markov sources with infinite state spaces. They provide an optimum threshold by using the Whittle's index policy. Again, for the pull-based model, similar system models are investigated for various age-related metrics \cite{akar2023optimum,champati2022detecting,bastopcu2021}.  In \cite{champati2022detecting} and \cite{bastopcu2021}, the sampling rates are allowed to change according to the latest estimation, similar to this paper. While binary freshness is used as the freshness metric in \cite{bastopcu2021}, an optimum policy to minimize age-penalty (defined as the time after the latest state transition) is derived in \cite{champati2022detecting}. In \cite{akar2023optimum}, a remote monitor samples multiple CTMC sources under a sampling constraint in continuous time, and closed-form expressions for optimum sampling policies are derived for three freshness metrics which are fresh when equal (FWE), fresh when sampled (FWS), and fresh when close (FWC). Another recent work in a continuous time setting is \cite{inoue2019aoi}, where the probability distribution of AoI, and the joint probability distribution of the observed process $X(t)$ and its estimation $\hat{X}(t)$, are derived.

\begin{figure}[t]
    \centering
    \includegraphics[width=0.75\columnwidth]{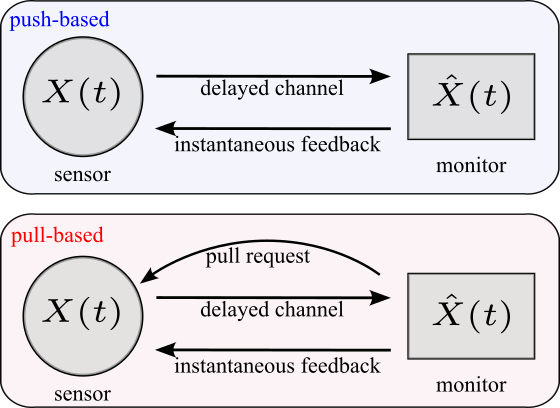}
    \caption{Push-based and pull-based models for a sensor and a monitor.}
    \label{fig:sys_pullpush}
    \vspace*{-0.4cm}
\end{figure}

In the pull-based model, the sensor starts transmitting information on the observed CTMC only when a pull request from the monitor is received. In particular, the monitor is allowed to make pull requests according to a Poisson process with intensity dependent of the current estimation. In the push-based scenario, the sensor, being aware of the AoII process, samples and transmits when the AoII process exceeds a random threshold (Erlang-$k$ distributed), which depends on the current value of the estimator. Transmission times are random (exponentially distributed) in both models allowing us to study the impact of network delays on the AoII.

The proposed analytical models for both push- and pull-based scenarios are based on the construction of a DTMC making state transitions at the embedded epochs of synchronization points, using the theory of absorbing CTMCs, and in particular the \emph{phase-type distribution} which stands for the distribution until absorption in a CTMC with one absorbing state \cite{latouche1999introduction}. The analytical models developed in this paper are based on the well-known properties of phase-type distributions which have also been used for information freshness and networked control problems in several existing works. In \cite{akar2023distribution}, the distribution of AoI and peak AoI metrics are derived by making use of phase-type distributions to deal with CTMCs with absorbing states. In another work \cite{scheuvens2021state}, the phase-type distribution is used to find the expected time before a certain number of consecutive packet failures happens in a wireless closed-loop control system.

To summarize the contributions of this paper, i) we develop a novel and computationally efficient analytical model for mean AoII in push- and pull-based sampling of CTMC processes, ii) the proposed analytical model relies on the well-known theory of absorbing CTMCs and can easily be extended to other settings including discrete time and other information freshness metrics, iii) it portrays a framework by which optimum system parameters can be selected.

\section{Preliminaries}\label{sec:prel}

The following are preliminaries on absorbing Markov chains that are needed to follow the proposed analytical method. Consider a CTMC $Y(t)$ with $K_1$ transient and $K_2$ absorbing states. Let the generator of this CTMC be written as,
\begin{align}
 \left[ \begin{array}{c|c}
   \bm A & \bm B \\
   \midrule
   \bm{0} & \bm{0} \\
\end{array}\right] \label{eq:Tmatrix}
\end{align}
where $\bm A_{K_1 \times K_1}$ and $\bm B_{K_1 \times K_2}$ correspond to the transition rates among the transient states, and from the transient states to the absorbing states, respectively. Upon merging all the absorbing states into one, the distribution of time until absorption, denoted by $T$, is known as the phase-type distribution \cite{latouche1999introduction} whose probability density function (PDF) is written in terms of the matrix exponential function of $t\bm A$ as,
\begin{align}
    f_T(t)=\bm{\beta} \mathrm{e}^{t\bm A}\bm{1},
    \label{eq:pdf}
\end{align}
from which the first and second moments can be obtained,
\begin{align}
    \mathbb{E}[T]=-\bm{\beta} \bm A^{-1} \bm{1}, \quad
    \mathbb{E}[T^2]=2\bm{\beta} \bm A^{-2} \bm{1}, \label{eq:mom12}
\end{align}
where $\bm{\beta} = \{ \beta_i \}$ is a ${1 \times K_1}$ row vector with $\beta_i$ standing for the initial probability in being transient state $i$, and $\bm{1}$ denotes a column vector of ones of appropriate size. Moreover, the probability of being absorbed in absorbing state $j$ is,
\begin{align}
    p_{j} & =-\bm{\beta} \bm{A}^{-1} \bm B \bm{e}^{(j)}, \label{eq:prob}
\end{align}
where $\bm{e}^{(j)}$ is a column vector of zeros of appropriate size except for its $j$th element which is one.

The absorbing CTMC whose generator is given in \eqref{eq:Tmatrix} can be converted to a DTMC with probability transition matrix as, 
\begin{align}
 \left[ \begin{array}{c|c}
   \bm D & \bm E \\
   \midrule
   \bm{0} & \bm{0} \\ 
\end{array}\right] \label{eq:TmatrixDiscrete1}
\end{align}
at the embedded epochs of state transitions. In particular, $\bm D = \{ d_{ij} \}$ can be written as, 
\begin{align}
    d_{ij}=\begin{cases}
        -\dfrac{a_{ij}}{a_{jj}}, & \text{if } i\neq j, \\ 0, & \text{otherwise.}
    \end{cases} \label{eq:diag}
\end{align}
A basic property about an absorbing Markov chain is that the expected number of visits to a transient state $j$ starting from a transient state $i$ is given by the $(i,j)$th entry of the so-called fundamental matrix $\bm F=(\bm I-\bm D)^{-1}$ \cite{kemeny1960finite}, where $\bm I$ denotes the identity matrix of suitable size.

\section{System Model}

Consider a sensor observing a finite state-space CTMC process $X(t) \in \mathcal{N}=\{1,2,\ldots,N\}$ with generator matrix $\bm Q = \{ q_{ij} \}$, where the state holding time at state $i$, denoted by $H_i$, is exponentially distributed with parameter $\sigma_i = -q_{ii}$, i.e., $H_i \sim \text{Exp}(\sigma_i)$, after which a state transition takes place to state $j$ with transition probability  $p_{ij}=-{q_{ij}}/{q_{ii}}$, for $i\neq j$. The monitor uses the last received update as its estimation, and the mismatch between the observed process and its estimation is measured with the AoII process, which is expressed as,
\begin{align}
\mbox{AoII}(t)=t-\max(\tau \; | \; X(\tau)=\hat{X}(\tau),\; \tau \leq t).
\end{align} 
We separately consider the {push-based} and {pull-based} transmission models which are illustrated in Fig.~\ref{fig:sys_pullpush}. 

In the push-based model, the sensor decides when to start transmitting its observation to the monitor. We assume a channel with exponentially distributed service time with parameter $\mu$ to account for network delays. However, we assume instantaneous feedback from the monitor and therefore the sensor is always aware of the remote estimator and also of the AoII process maintained at the monitor. The sensor then waits for the AoII to exceed a random threshold $\Gamma_i$ which depends on the current value $i$ of the estimator. In this paper, we focus on Erlang-$k$ distributed thresholds where an Erlang-$k$ random variable is the sum of $k$ exponential random variables with the same parameter whose variance approaches zero as $k \rightarrow \infty$, i.e., deterministic distribution. Clearly, the exponential distribution is a sub-case of Erlang-$k$ when $k$=1. 

In the pull-based model, sampling is initiated with a pull request from the monitor which sends these requests according to a Poisson process with intensity depending on the current value of the estimator. Pull requests from the monitor reach the sensor immediately, similar to instantaneous feedback. In both models, we assume that the sensor preempts its transmission if the observation changes before the transmission is over. After the preemption, while a new transmission is always initiated in the push-based model, a new pull request is required for initiating a transmission in the pull-based model. Additionally, a constraint on the sampling rate is placed, so that the sensor cannot transmit the observations constantly. We leave two-way network delays, more advanced channel models (such as queued models), and deterministic thresholds for future work.

\section{Analytical Method}

We first need the following definition: A time point $t$ is a synchronization point (SP) at value $S_i$ if $X(t)$ and $\hat{X}(t)$ are just synchronized at the value $i$, i.e., $X(t)=\hat{X}(t)=i$, $X(t^-) \neq \hat{X}(t^-)$. The interval between the SP $t$ at value $S_i$ and the next SP is called cycle-$i$. Subsequently, we consider a DTMC embedded at the epochs of synchronization points whose states are the synchronization values $S_i, i \in \mathcal{N}$. While the transition from $S_i$ (at the current SP) to $S_j$ (at the next SP) for $j \neq i$ is incurred with a timely reception of a sample, a self-transition from $S_i$ to itself can also occur when $X(t)$ returns to its initial value $i$ without a new sampling operation. Let $p_{ij}$ and $\pi_i$ denote the transition probability from synchronization value $S_i$ to $S_j$, and the steady-state probability of being in state $S_i$ at an SP. The steady-state vector $\boldsymbol{\pi} = \{ \pi_i\}$ satisfies
\begin{align}
    \bm{\pi}&=\bm{\pi} \bm{P}, \quad \bm{P} \bm{1}=\bm{1}, 
\end{align}
where $\bm P=\{ p_{ij} \}$. Then, we obtain $\bm{\pi}$ as,
\begin{align}
    \bm{\pi } &=\bm{1}^T(\bm{P}+\bm{1}\bm{1}^T-\bm{I})^{-1}. \label{eq: oneonetranspose}
\end{align}
We note that the actual values of the transition probabilities will be dependent on the model (push- or pull-based) and the particular sampling policy of interest. 

Let $a_i$, $d_i$, and $r_i$, denote the area under the AoII curve, duration, and average number of samples taken, respectively, of cycle-$i$. Let $\mbox{AoII}$ and $R$ denote the average AoII and the average sampling rate of the overall system, respectively. Then,
\begin{align}
   \mbox{AoII} & =\dfrac{\sum_{n=1}^{N}\pi_n a_n}{\sum_{n=1}^{N} \pi_n d_n}, 
    \quad R=\dfrac{\sum_{n=1}^{N}\pi_n r_n}{\sum_{n=1}^{N} \pi_n d_n}. \label{eq:age}
\end{align}
In the next two subsections, for push- and pull-based sampling, respectively, and for a given sampling policy, we will first derive the transition matrix $\bm{P}$ of the embedded DTMC, from which we will obtain $\bm{\pi}$, and the quantities $a_i$, $d_i$, and $r_i$, $i \in \mathcal{N}$ which will allow us to obtain $\mbox{AoII}$ and $R$ via \eqref{eq:age}.

\subsection{Push-Based Sampling}

\begin{figure}[t]
    \centering
   \includegraphics[width=0.9\columnwidth]{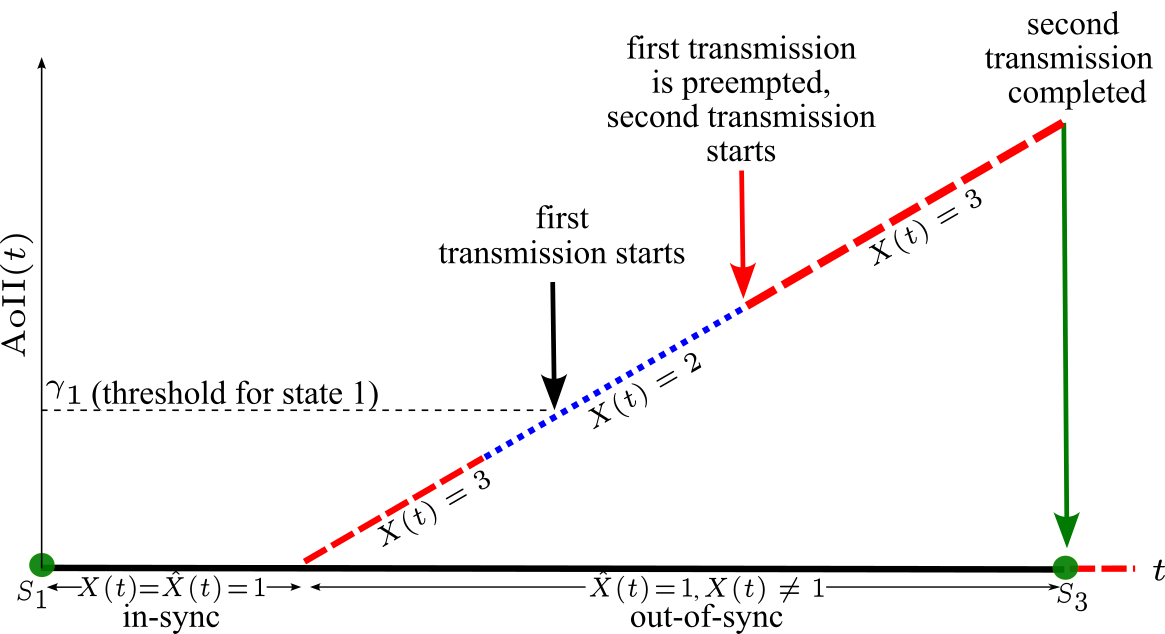} 
    \caption{A sample path for $X(t)$, $\hat{X}(t)$ and $\text{AoII}(t)$ for push-based transmission for an example scenario. Green circles denote the synchronization points.}
    \label{fig:trans2}
    \vspace*{-0.4cm}
\end{figure}

A sample path for the push-based sampling model is illustrated in Fig.~\ref{fig:trans2} for $N=3$ starting from $S_1$. Let $\gamma_1$ be a realization of the random threshold $\Gamma_1$. When $\mbox{AoII}(t)$  reaches $\gamma_1$, the sensor samples $X(t)$ at state 2, and starts transmitting this observation which requires an exponentially distributed service time with parameter $\mu$. However, before the transmission is complete, a state transition from state 2 to state 3 takes place. Therefore, the sensor preempts the previous transmission and re-samples $X(t)$ at state 3. Finally, the system reaches the next synchronization value $S_3$ at which $X(t)$ and $\hat{X}(t)$ are synchronized at 3, since this second transmission has completed without encountering a state change in the original process. It is also possible that we can reach the next synchronization value $S_1$ by a state change of the original process $X(t)$ from states 2 or 3 to the original state 1, but this situation is not illustrated in Fig.~\ref{fig:trans2}.

\begin{table}[t]
\caption{Transition rates for the absorbing CTMC process $Y_i(t)$ for push-based sampling.}
\centering
\begin{tabular}{|c|c|c|} 
 \hline
 \multicolumn{3}{|c|}{Transition Rates from $(j,l)$ for $Y_i(t)$} \\
 \hline
 To & Condition & Value \\ 
 \hline
$(j,l+1)$ & $ l \neq k $ & $k{\theta_i}$ \\
 \hline
$(j',l)$ & $ j' \neq j, \ \forall l $ & $q_{jj'}$ \\
 \hline
$S_i$ & $ \forall l$ & $q_{ji}$ \\
\hline
$S_j$ & $l=k$ & $\mu$ \\
\hline 
\end{tabular}
\label{tab:Table1}
\vspace*{-0.4cm}
\end{table}

In the following, we focus on the particular synchronization value $S_i, \; i\in \mathcal{N}$ and describe the analytical method to find $p_{ij}, \ j \in \mathcal{N}$, $a_i$, $b_i$, and $r_i$ when $\Gamma_i \sim \text{Erlang-$k$}(\theta_i)$. The procedure needs to be repeated  $\forall S_j, \ j \neq i$. Let us revisit Fig.~\ref{fig:trans2} and concentrate on the interval when $\text{AoII}(t) >0$ until $\text{AoII}(t)$ becomes zero. We will now show that this interval can be modeled as a phase-type distribution described in Section~\ref{sec:prel} with $Y_i(t)$ representing the corresponding absorbing CTMC process. For this purpose, we define a transient state as the pair $(j,l)$, where $j \neq i$ is the actual state of $X(t)$ and $l \in \{ 0, \dots, k-1\}$ corresponds to the stage of the Erlang-$k$ random variable and $l=k$ represents the situation that the threshold value is exceeded. In state $(j,l)$ for $l<k$, the system is out-of-sync but the threshold is not reached yet, and the absorbing CTMC $Y_i(t)$ starts operation from state $(j,0)$ with probability $\frac{q_{ij}}{\sigma_i}$. In state $(j,k)$, the system is still out-of-sync but the threshold is reached and therefore there is an ongoing transmission. Finally, we have the absorbing states $S_j, \ j\in \mathcal{N}$. According to the notation of Section~\ref{sec:prel}, we have $K_1=(k+1)(N-1)$ transient states and $K_2=N-1$ absorbing states. Then, the generator for $Y_i(t)$ is in the form \eqref{eq:Tmatrix} with the matrices $\bm A$, $\bm{B}$, and $\bm{\beta}$ of \eqref{eq:Tmatrix} replaced with  $\bm A^{(i)}$, $\bm{B}^{(i)}$, and $\bm{\beta}^{(i)}$, respectively, for which the characterizing matrix $\bm A^{(i)}$ can be written as,
\begin{align}   
    \!\!\!\begin{bmatrix}
        \bm{Q}^{(-i)}\!-\!k\theta_iI & k\theta_iI & \bm{0} &\dots & \bm{0} \\
        \bm{0} & \bm{Q}^{(-i)}\!-\!k\theta_iI & k\theta_iI & \dots & \bm{0} \\
        \vdots & & & & \vdots \\
        \bm{0} & \dots &  &\bm{0} & \bm{Q}^{(-i)}\!-\!\mu \bm{I}
    \end{bmatrix}\!\!
\end{align}
and the other two matrices $\bm{B}^{(i)}$ and $\bm{\beta}^{(i)}$ can be written as,
\begin{align}
 \bm{B}=\begin{bmatrix}
        \bm{q}^{(i)} & \bm{0} \\
        \vdots & \vdots \\
        \bm{q}^{(i)} & \mu \bm{I}
    \end{bmatrix}, \quad 
    \bm{\beta}^{(i)} = \begin{bmatrix}
        \dfrac{1}{\sigma_i}\bm{q}_r^{(i)} & \smash[b]{  \underbrace{ \begin{matrix} 0 &  \cdots &  0 \end{matrix}}_{k(N-1)}},
    \end{bmatrix}
\end{align}
where $\bm{Q}^{(-i)}$ is $\bm{Q}$ but its $i$th row and $i$th column removed, and $\bm{q}^{(i)}$ is the $i$th column of $\bm{Q}$ excluding $q_{ii}$, and  $\bm{q}_r^{(i)}$ is the $i$th row of $\bm{Q}$ excluding $q_{ii}$.

We observe that the duration before absorption in this model, denoted by $T^{(i)}$, is equal to the duration of the out-of-sync interval in cycle-$i$. Since $T^{(i)}$ is phase-type distributed, the area under the AoII curve $a_i$ and the duration $d_i$ of cycle-$i$ using \eqref{eq:mom12} are,
\begin{align}
    a_i & =\mathbb{E}\left[\frac{(T^{(i)})^2}{2}\right] = \bm{\beta}^{(i)} (\bm A^{(i)})^{-2} \bm{1}, \label{eq:mom2_exp} \\
    d_i& =\mathbb{E}[T^{(i)}+H_i]= -\bm{\beta}^{(i)}(\bm A^{(i)})^{-1} \bm{1}+\dfrac{1}{\sigma_i}. \label{eq:mom1_exp}    
\end{align}
Finally, we can calculate from \eqref{eq:prob} the transition probability from the synchronization value $S_i$ to $S_j$, $p_{ij}$, as
\begin{align}
    p_{ij}&=-\bm{\beta}^{(i)} (\bm A^{(i)})^{-1} \bm{B}^{(i)} \bm{e}^{(j)}. \label{eq:prob_exp}
\end{align}
In order to obtain $r_i$, we need to find the average number of times a transient state $(j,k)$ is visited before absorption. For this purpose, we first obtain the absorbing MC in discrete time with the probability transition matrix in the form 
\begin{align}
 \left[ \begin{array}{c|c}
   \bm D^{(i)} & E^{(i)}\\
   \midrule
   \bm{0} & \bm{0} \\
\end{array}\right] \label{eq:TmatrixDiscrete2}
\end{align}
with $\bm D^{(i)}$ obtained from $\bm A^{(i)}$ as in \eqref{eq:diag}, resulting in the fundamental matrix  $\bm{F}^{(i)}$,
\begin{align}
    \bm{F}^{(i)} & =(\bm{I}-\bm D^{(i)})^{-1}.
\end{align}
Any time a state $(j,k)$ is visited, a new sample is taken. Therefore, following Section~\ref{sec:prel}, we can write $c_i$ as the average number of visits to any of the states $(j,k)$ as,
\begin{align}
    r_i=\bm{\beta}^{(i)} \bm{F}^{(i)} \bm{f}, \label{eq:cost_exp}
\end{align}
where $\bm{f}$ is a column vector whose last $N-1$ elements are one, and its remaining elements are zero.  This concludes the calculations required for cycle-$i$. We repeat this procedure for all cycles to obtain the discrete time probability transition matrix $\bm{P}$ using \eqref{eq:prob_exp} and its steady-state vector $\bm{\pi}$ using \eqref{eq: oneonetranspose}, which subsequently allow us to obtain AoII and the average sampling rate $R$ through the expressions in \eqref{eq:age}, completing the analysis for $\Gamma_i \sim \text{Erlang-}k(\theta_i)$.  

\subsection{Pull-Based Sampling}

In this scheme, the monitor sends pull requests according to a Poisson process with rate $\lambda_i$ when the last estimation is $\hat{X}(t)=i$, and the sensor transmits the sample with an exponentially distributed service time with parameter $\mu$. Different from the push-based model, if the observed process $X(t)$ changes before the transmission is completed and the sensor preempts the transmission, the sensor does not start another transmission until it receives another pull request. Further, in this model, a transmission may also happen while the source and the monitor are in-sync. However, this redundant sample does not affect the AoII and the estimation maintained at the monitor, but increases the average sampling rate.

\begin{figure}[t]
    \centering
    \includegraphics[width=0.95\columnwidth]{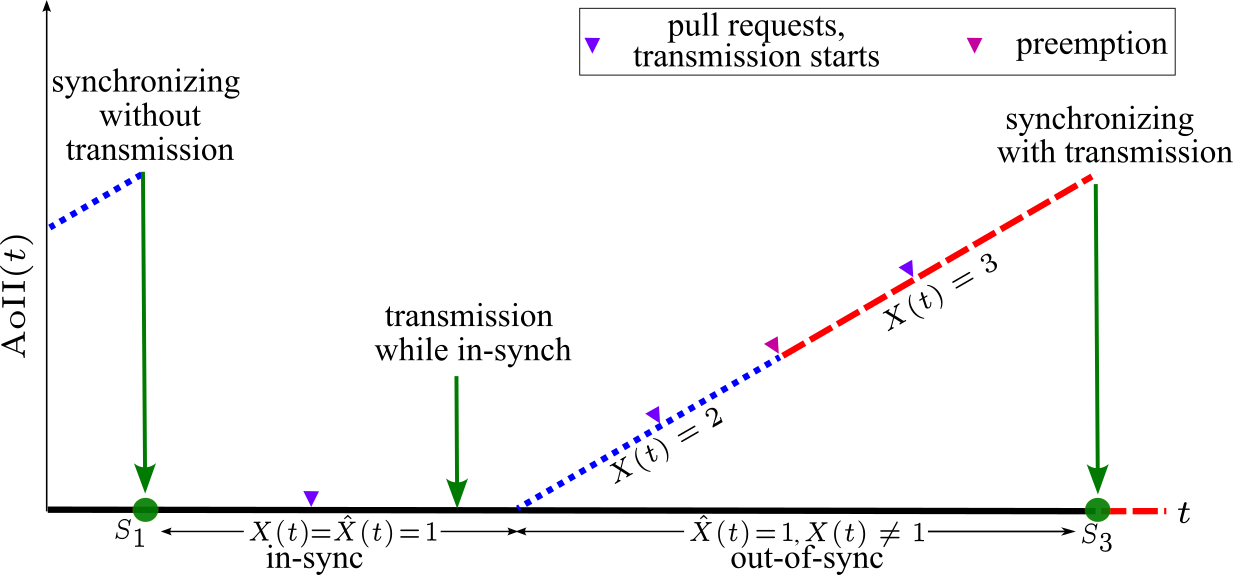} 
    \caption{A sample path for $X(t)$, $\hat{X}(t)$ and $\text{AoII}(t)$ for pull-based transmission for an example scenario.}
    \label{fig:trans}
    \vspace*{-0.4cm}
\end{figure}

\begin{table}[t]
\caption{Transition rates for the absorbing CTMC process $Y_i(t)$ for pull-based sampling.}
\centering
\begin{tabular}{|c|c|c|} 
 \hline
 \multicolumn{3}{|c|}{Transition Rates from $(j,l)$ for $Y_i(t)$} \\
 \hline
 To & Condition & Value \\ 
 \hline
 $(j,1)$ & $l=0 $ & ${\lambda_i}$ \\
 \hline
 $(j',0)$ & $ j' \neq j, \ l = 0,1$ & $q_{jj'}$ \\
 \hline
 $S_i$ & $ l = 0,1 $ & $q_{ji}$ \\
\hline
 $S_j$ & $l=1$ & $\mu$ \\
\hline 
\end{tabular}
\label{tab:Table2}
\vspace*{-0.4cm}
\end{table}

A sample path of this transmission model is illustrated in Fig.~\ref{fig:trans} for an example scenario with $N=3$, while pull-requests are illustrated with blue arrowheads, and preemption and completion times of transmissions are shown with purple and green arrowheads, respectively. At the beginning, the synchronization value $S_1$ is reached when $X(t)$ changes from state 2 to state 1, which matches with the estimation at this SP. While in-sync, a pull request is received upon which a transmission starts which is eventually completed. However, this particular transmission does not have an effect on the evolution of AoII. Synchronization is broken after $X(t)$ visits state 2 upon which two further pull requests are received. While the first transmission is preempted because of the state change, the transmission following the second pull request is completed at which the synchronization value $S_3$ is reached.

Similar to the push-based model, we introduce the absorbing CTMC process $Y_i(t)$ corresponding to the out-of-sync interval in cycle-$i$. The CTMC $Y_i(t)$ has absorbing states $S_i, i\in \mathcal{N}$ and transient states as the pair $(j,l)$ where $j \neq i$ corresponds to the state of $X(t)$, and $l$ is a binary state $\{0,1\}$. In state $(j,0)$, the sensor waits for a pull request to start transmission and $Y_i(t)$ starts operation from this state with probability $\frac{q_{ij}}{\sigma_i}$, and state $(j,1)$ indicates an ongoing transmission. All transition rates are summarized in Table~\ref{tab:Table2}. By using them, we can express the characterizing matrices $\bm A^{(i)}$, $\bm{B}^{(i)}$ and $\bm{\beta}^{(i)}$ of the absorbing CTMC $Y_i(t)$ as in Section~\ref{sec:prel},
\begin{align}
\bm A^{(i)}&=\begin{bmatrix}
    \bm{Q}^{(-i)}-\lambda_i \bm{I} & \lambda_i \bm{I} \\
    \bm{Q}^{(-i)}-\mbox{diag}(\bm{Q}^{(-i)}) & \mbox{diag}(\bm{Q}^{(-i)})-\mu \bm{I} 
\end{bmatrix}, \\
\bm{B}_{(i)}&=\begin{bmatrix}
    \bm{q}_{(i)}& \bm{0} \\ \bm{q}_{(i)} & \mu \bm{I}
\end{bmatrix}, \        
\bm{\beta}^{(i)} = \begin{bmatrix}
    \dfrac{1}{\sigma_i}\bm{q}_r^{(i)} & \smash[b]{  \underbrace{ \begin{matrix} 0 &  \cdots &  0 \end{matrix}}_{N-1}} 
\end{bmatrix},
\end{align}
where $\mbox{diag}(\bm M)$ is a diagonal matrix composed of the diagonal elements of $\bm M$. Again, $a_i$, $d_i$, $p_{ij}$ are obtained using \eqref{eq:mom2_exp}, \eqref{eq:mom1_exp} and \eqref{eq:prob_exp}, respectively. However, in order to calculate $r_i$, we also should consider the number of transmissions while in-sync, which is equal to the average number of pull requests before synchronization is broken. We can model that by considering an absorbing CTMC ${Z}_i(t)$ with two transient states and one absorbing state. In the first state (which we initially start the operation at), we wait for a pull request with rate $\lambda_i$ whereas we complete transmission with rate $\mu$ in the second state. On the other hand, the absorbing state is reached upon a state change in the original process which occurs with rate $\sigma_i$. Therefore, the characterizing matrices $\hat{\bm{A}}^{(i)}$ and $\hat{\bm D}^{(i)}$ of the process $Z_i(t)$ given in Section~\ref{sec:prel} can be expressed as, 
\begin{align}
\hat{\bm{A}}^{(i)}=\begin{bmatrix}
       -\sigma_i-\lambda_i & \lambda_i \\  \mu &-\mu-\sigma_i 
    \end{bmatrix}, \     
    \hat{\bm D}^{(i)}=\begin{bmatrix}
       0 & \frac{\lambda_i}{\lambda_i+\sigma_i} \\ \frac{\mu}{\mu+\sigma_i} & 0
    \end{bmatrix},
\end{align}
which results in the following expression for $r_i$,
\begin{align}
r_i&=\bm{\beta}^{(i)} (\bm{I}-\bm D^{(i)})^{-1} \bm{f}+ \begin{bmatrix}
        1 & 0
    \end{bmatrix} (\bm{I}-\hat{\bm {D}}^{(i)})^{-1} \begin{bmatrix}
        0 \\ 1
    \end{bmatrix}, \\
    &=\boldsymbol{\beta}^{(i)} (\bm{I}-\bm D^{(i)})^{-1} \bm{f}+ \dfrac{\lambda_i(\mu+\sigma_i)}{\lambda_i\sigma_i+\sigma_i\mu+\sigma_i^2},
\end{align}
where $\bm{f}$ is a column vector whose last $N-1$ elements are one, and its remaining elements are zero.
It is worth noting that the average sampling rate may be smaller than the overall pull request rate since the sensor ignores  pull requests that arrive during an ongoing transmission.

\section{Numerical Results}

\begin{figure}[t]
    \begin{center}
    \subfigure[]{\includegraphics[width=0.475\columnwidth]{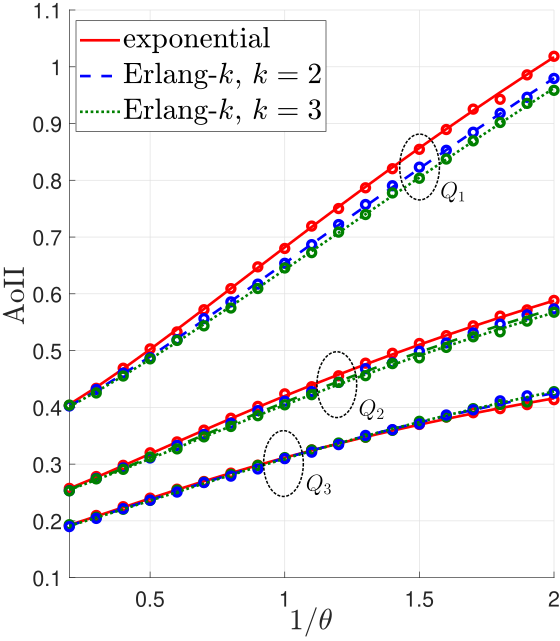}} ~ 
    \subfigure[]{\includegraphics[width=0.475\columnwidth]{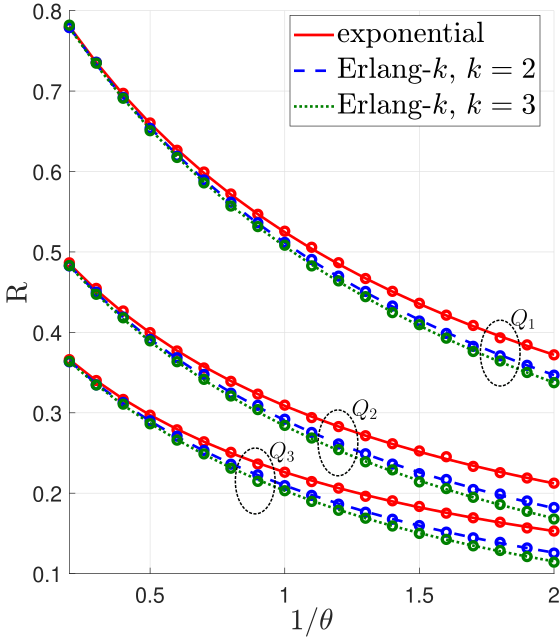}}  
    \end{center}  
    \vspace*{-0.4cm}
    \caption{a) AoII,  b) average sampling rate $R$ of the push-based transmission model as a function of $\theta$, for three CTMC processes and for three different values of the parameter $k$. While lines are for the analytical results, circles correspond to the simulation results.}
    \label{fig:sim_push}
    \vspace*{-0.8cm}
\end{figure}

In the first numerical example, we verify the push-based analytical model with simulation results for three different scenarios. For all of these simulations, we select the channel transmission rate as $\mu=1$ and $\mathbb{E}[\Gamma_i] = \theta$. Three source processes are studied with the following generators,
\begin{align}
\bm{Q}_1 & = \{ q_{1,ij} \}_{5\times5}, \quad q_{1,ij} = 0.25, \ j \neq i, \quad q_{1,ii} = -1, \label{source1} \\
    \bm{Q}_2&= \begin{bmatrix}
        -1 & 0.7 & 0.3 \\
        0.2 & -0.6 & 0.4 \\
        0.1 & -0.7 & 0.8
    \end{bmatrix}, \ \bm{Q}_3= \begin{bmatrix}
        -0.5 & 0.5 \\ 0.75 & -0.75
    \end{bmatrix}, \label{sources23}
\end{align}
where the first one has homogeneous states and the other two having heterogeneous states. In Fig.~\ref{fig:sim_push}, AoII and average sampling rate $R$ are depicted as a function of $\theta^{-1}$ for all the three information sources for Erlang-$k$ distributed thresholds for $k=1\text{ (i.e., exponentially distributed case), }2,3$ using both simulations and the analytical model. We observe an exact agreement between simulations and analytical results for all the cases. When $\theta$ increases, we attain lower sampling rates at the expense of an increase in AoII. Additionally, we observe that when the order of the Erlang-$k$ distribution increases, i.e., the variance of the threshold decreases, then a lower AoII is obtained at a lower average sampling rate. Studying Erlang-$k$ distribution with larger $k$ or the deterministic distribution, for improved AoII performance is left for future work.

\begin{figure}[t]
    \centering
    \includegraphics[width=0.8\columnwidth]{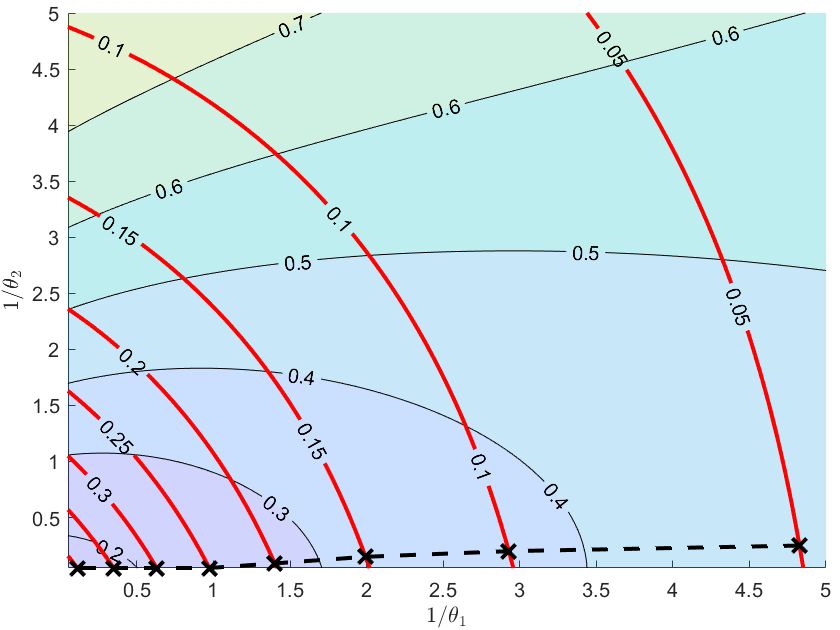}
    \vspace*{-0.1cm}
    \caption{Contour maps of AoII and $R$ with respect to the sampling policy $(\theta_1,\theta_2)$. The points which result in the minimum AoII values among the policies with the same average sampling rate pairs are marked with a cross.}
    \label{fig:push_2d}
\end{figure}

Next, we focus on the information source with generator $\bm{Q}_3$ with $N=2$, $\mathbb{E}[\Gamma_i] = \theta_i, i=1,2,$ and employ $k=3$ while varying $\theta_1$ and $\theta_2$. In Fig.~\ref{fig:push_2d}, we study the performance of the sampling policies characterized with the pair $(\theta_1,\theta_2)$, using a contour map. The red contour lines (resp. gray contour lines) correspond to the set of policies with the same average sampling rate (resp. same AoII). Additionally, for a given sampling rate budget, the policy that results in the minimum AoII is marked by a cross. We conclude that, if there is sufficient sampling rate budget, the best policy is to choose a small threshold for the faster state (state 2 with a lower state holding time), and use the rest of the budget for the slower state (state 1), and the best strategy converges to the $(0,0)$ policy while the sampling rate budget increases.

\begin{figure}[t]
    \begin{center}
    \subfigure[]{\includegraphics[width=0.475\columnwidth]{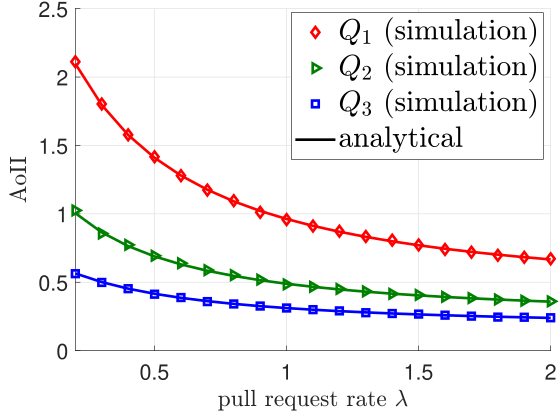}} ~ 
    \subfigure[]{\includegraphics[width=0.475\columnwidth]{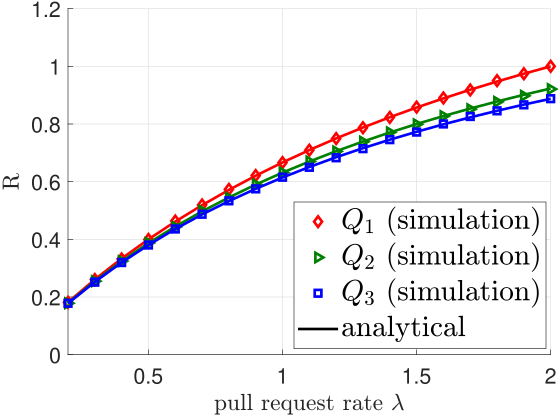}}
    \end{center}
    \vspace*{-0.4cm}
    \caption{a) AoII, b) average sampling rate $R$ of the pull-based transmission model as a function of the common pull request rate $\lambda$.}  
    \label{fig:sim_pull}
    \vspace*{-0.6cm}
\end{figure}

\begin{figure}[t]
    \centering
    \includegraphics[width=0.8\columnwidth]{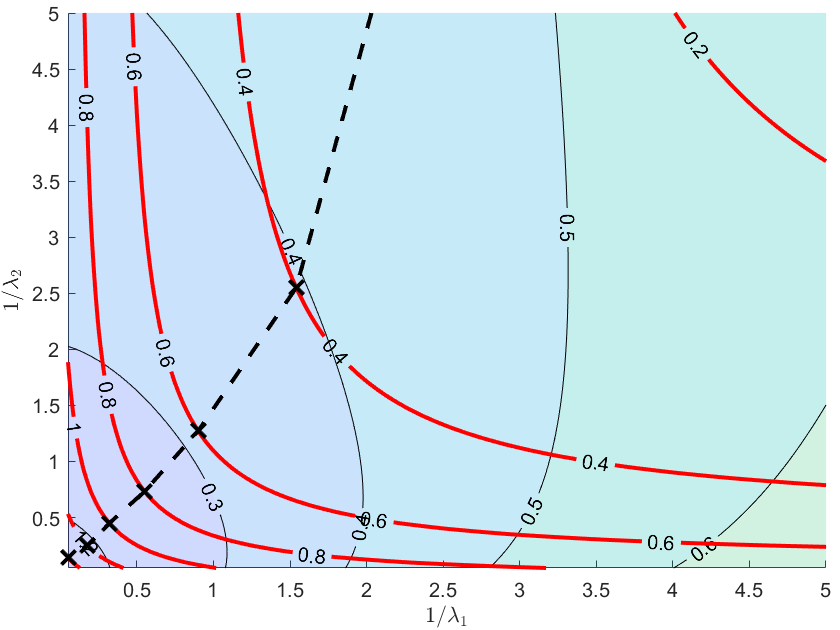}
    \vspace*{-0.1cm}
    \caption{Contour maps of AoII and $R$ with respect to the sampling policy $(\lambda_1,\lambda_2)$. The points which result in the minimum AoII values among the policies with the same average sampling rate pairs are marked with a cross.}    
    \label{fig:pull2d}
    \vspace*{-0.4cm}
\end{figure}

In the second numerical example, similar simulations are repeated for the pull-based model with the same generators given in \eqref{source1} and \eqref{sources23} and $\mu=1$. In Fig.~\ref{fig:sim_pull}, AoII is depicted as a function of pull request rate $\lambda=\lambda_i, \ i\in \mathcal{N}$ both with simulations and the analytical model, and there is perfect agreement between the two curves. Moreover, we observe that AoII decreases and $R$ increases monotonically when $\lambda$ is increased. Additionally, while the rate of pull requests are lower than the transmission rate $\mu=1$, almost all pull requests lead to a sampling operation resulting in a linear relation between them. As the rate of pull requests increases, some of these requests arrive when a transmission is ongoing and they are therefore ignored. In Fig.~\ref{fig:pull2d}, we study the performance of the sampling policies for the pull-based model characterized with the pair $(\lambda_1,\lambda_2)$, using a contour map using the same convention as in Fig.~\ref{fig:push_2d}. Unlike the push-based scenario, we observe that frequent sampling while the estimation is the slow state may not lead to the best policy under a given sampling rate budget. We observe that a more balanced sampling policy (note the positions of the cross markers in Fig.~\ref{fig:pull2d}) is more well-suited for the pull-based model. The reason behind this is that while the push-based system only samples when the estimation is out-of-sync, the pull-based system may take samples even when in-sync, which increases the average sampling rate without impacting the AoII performance.

\section{Conclusions and Future Work}

We studied the average AoII and the average sampling rate of push- and pull-based sampling systems when the information source is a CTMC process. We developed novel and computationally efficient analytical models for both systems using a common framework based on the theory of absorbing CTMCs. For the push-based system and using the analytical models, we have observed that both AoII and average sampling rate performances are improved when the variance of the sampling threshold decreases. Moreover, the optimum sampling policy (leading to minimum AoII under a sampling budget) leads to lower sampling thresholds for slower states. We have also observed that this situation is very different in pull-based systems for which we have observed that more balanced pull requests lead to optimum sampling. Future work will consist of two-way delays, other sampling policies including deterministic thresholds in push-based systems, and the use of MDPs in deriving the optimum policies. Using reinforcement learning for CTMC sources whose dynamics are not known in advance may be an interesting direction.

\bibliographystyle{unsrt}
\bibliography{bibl}
\end{document}